\documentclass[fleqn]{article}
\usepackage{amsmath}
\usepackage{amssymb}
\usepackage{amsthm,easymat}
\usepackage[nice]{nicefrac}

\usepackage{ifpdf}
\ifpdf
\usepackage[pdftex]{graphicx}
\else
\usepackage[dvips]{graphicx}
\fi

\ifpdf
\usepackage[pdftex,colorlinks,linkcolor=blue,citecolor=blue]{hyperref}
\else
\usepackage{hyperref}
\fi

\begin{document}
\author{Uriel Feige \and Eran Ofek}
\title{Random 3CNF formulas elude the Lov\'{a}sz theta function}

\maketitle

\newtheorem{thm}{Theorem}[section]
\newtheorem{corollary}[thm]{Corollary}
\newtheorem{lemma}[thm]{Lemma}
\newtheorem{theorem}[thm]{Theorem}
\newtheorem{prop}[thm]{Proposition}

\newtheorem{conjecture}[thm]{Conjecture}
\newtheorem{definition}[thm]{Definition}

\newcommand{\bits}{\{ 0,1 \}}
\newcommand{\e}{\epsilon}
\newcommand{\norm}[2]{\| #2 \|_{_{#1}}}
\newcommand{\equivalent}{\stackrel{\rm [1]}{\approx}}
\newcommand{\eqdef}{\stackrel{\vartriangle}{=}}
\newcommand{\Ex}[2]{\underset{#1}{E} [#2] }
\newcommand{\discup}{$\bigcup$ \hspace{-2.7ex} $\cdot$ \hspace{0.5ex} }
\newcommand{\twolines}[2]{\stackrel{#1}{#2}}
\newcommand{\pfrac}[2]{\left(\frac{#1}{#2}\right)}
\newcommand{\vs}{} 
\newcommand{\svs}{} 

\def\draft{1}   

\ifnum\draft=1 
    \def\ShowAuthNotes{1}
\else
    \def\ShowAuthNotes{0}
\fi

\ifnum\ShowAuthNotes=1
  \newcommand{\authnote}[2]{{ \bf [#1's note: #2]}}
\else
  \newcommand{\authnote}[2]{}
\fi

\newcommand{\PrIS}{1 - 1/n}
\newcommand{\PrNotIS}{1/n}
\newcommand{\PrExpansion}{1 - 1/n}
\newcommand{\PrNotExpansion}{1/n}
\newcommand{\err}{e^{-\Omega(\e^2 d)}}
\newcommand{\inprod}[2]{\langle #1,#2 \rangle}

\begin{abstract}
Let $\phi$ be a 3CNF formula with $n$ variables and $m$ clauses. A
simple nonconstructive argument shows that when $m$ is
sufficiently large compared to $n$, most 3CNF formulas are not
satisfiable. It is an open question whether there is an efficient
refutation algorithm that for most such formulas proves that they
are not satisfiable. A possible approach to refute a formula
$\phi$ is: first, translate it into a graph $G_{\phi}$ using a
generic reduction from 3-SAT to max-IS, then bound the maximum
independent set of $G_{\phi}$ using the Lov\'{a}sz $\vartheta$
function. If the $\vartheta$ function returns a value $<m$, this
is a certificate for the unsatisfiability of $\phi$. We show that
for random formulas with $m<n^{3/2 -o(1)}$ clauses, the above
approach fails, i.e. the $\vartheta$ function is likely to return
a value of $m$.
\end{abstract}

\section{Introduction}\label{sec: introduction}

A 3CNF formula $\phi$ over $n$ variables is a set of $m$ clauses,
where each clause contains exactly $3$ literals. A formula $\phi$ is
satisfiable if there is an assignment to its $n$ variables that sets
at least one literal in every clause to "true". The 3-SAT problem of
deciding whether an input 3CNF formula $\phi$ is satisfiable is
NP-hard. In this paper we consider a certain heuristic for 3-SAT. A
heuristic for satisfiability may try to find a satisfying assignment
for an input formula $\phi$ if one exists. A refutation heuristic
may try to prove that no satisfying assignment exists.

How does one measure the quality of a refutation heuristic? A
possible test may be to check how good the heuristic is on a random
input. But then, how do we generate a random unsatisfiable formula?
To answer this question we review some known properties of random
3CNF formulas. The satisfiability property has the following
interesting threshold behavior. Let $\phi$ be a random 3CNF formula
with $n$ variables and $cn$ clauses (each new clause is chosen
independently and uniformly from the set of all possible clauses).
As the parameter $c$ governing the density of the formula is
increased, it becomes less likely that $\phi$ is satisfiable, as
there are more constraints to satisfy. In \cite{FriedgutBo99} it is
shown that there exists $c_n$ such that for $c < c_n(1-\e)$ almost
surely $\phi$ is satisfiable, and for $c > c_n(1+\e)$, $\phi$ is
almost surely unsatisfiable (for some $\e$ which tends to zero as
$n$ increases). It is also known that $3.52 < c_n  < 4.596$
\cite{KaporisKiLa03,HajiaghayiSo03,JansonStVa00} and it is widely
believed that $c_n$ converge to some constant $c$. We will use
random formulas with $cn$ clauses (for $c
> c_n(1 + \e)$) to measure the performance of a refutation
heuristic.
Notice that for any $n$, as $c$ is increased (for
$c>c_n(1+\e)$), the algorithmic problem of refutation becomes less
difficult since we can always ignore a fixed fraction of the
clauses.

In this paper we analyse a semidefinite programming based refutation
algorithm which was introduced at \cite{Feige02}, and show that for
random formulas of certain densities (well above the satisfiability
threshold) this algorithm fails.

The algorithm itself is simple to describe (to readers familiar
with some of the previous work).

\begin{enumerate}

\item Given an input 3CNF formula $\phi$, apply to it a standard
reduction from max 3-SAT to maximum independent set, resulting in
a graph $G_{\phi}$. The size of the maximum independent set in
$G_{\phi}$ is equal to the maximum number of clauses that can be
simultaneously satisfied in $\phi$.

\item
Compute the Lov\'{a}sz $\vartheta$ function of the graph
$G_{\phi}$. This provides an upper bound on the size of the
maximum independent set of $G_{\phi}$.

\item If $\vartheta(G_{\phi}) < m$, then output "unsatisfiable".
Otherwise return "do not know".

\end{enumerate}

We now describe the graph $G_{\phi}$ in more detail. Recall that
for a 3CNF clause, there are seven different assignments to its
three literals that satisfy the clause. For each clause of $\phi$
the graph contains a clique of $7$ vertices, which we call a
\emph{cloud}. Hence $G_{\phi}$ contains $7m$ vertices. Each vertex
of the clause cloud is associated with a different assignment to
the three literals of the clause that satisfies the clause.
Vertices of different clouds are connected by an edge if they are
associated with contradicting assignments. (Namely, if there is a
variable that is assigned to true by one of the assignments and to
false by the other. For the same reason, the vertices within a
cloud form a clique.)

The $\vartheta$ function of any graph $G$ upper bounds the maximum
independent set in it, and can be computed in polynomial time up
to arbitrary precision, using semidefinite programming. The fact
that the vertices of $G_{\phi}$ can be covered by $m$ cliques
implies that $\vartheta(G_{\phi}) \leq m$. Thus, if $\phi$ is
satisfiable then the value of the theta function will be exactly
$m$. If the value of the theta function is $< m$ then $\phi$ is
unsatisfiable.

The above algorithm has one sided error, in the sense that it will
never say ``unsatisfiable'' on a satisfiable formula, but for some
unsatisfiable formulas it will fail to output ``unsatisfiable''.
If for some formula $\phi$ the algorithm outputs `unsatisfiable',
then the algorithm execution on $\phi$ is a witness for the
unsatisfiability of $\phi$.

Our main result is that for random 3CNF formula $\phi$ with $m <
n^{3/2 -o(1)}$ clauses it is very likely that
$\vartheta(G_{\phi})=m$.

\subsection{Related work}

A possible approach for refuting a formula $\phi$ is to find a
resolution proof for the unsatisfiability of $\phi$. However,
Chvatal and Szemeredi \cite{ChvatalSz88} proved that a resolution
proof of a random 3CNF formula with linear number of clauses is
almost surely of exponential size. A result of a similar flavor for
denser formulas was given by Ben-Sasson and Wigderson
\cite{Ben-SassonWi01} who showed that a random formula with $n^{3/2
-\e}$ clauses almost surely requires a resolution proof of size
$2^{\Omega(n^{\e /(1-\e)})}$. These lower bounds imply that finding
a resolution proof for a random formula is computationally
inefficient.

A simple refutation algorithm can be used to refute random
instances with $cn^2$ clauses, when $c > 2/3$. This is done by
selecting all the clauses that contain a variable $x$. Fixing $x$
to be true leaves about half of the selected clauses as a random
2-cnf formula with roughly $3cn/2 > n$ clauses. This formula is
unlikely to be satisfiable, and its nonsatisfiability can be
verified by a polynomial time algorithm for 2SAT. The same can be
done when fixing $x$ to be false.

A spectral approach introduced by Goerdt and
Krivelevich~\cite{GoerdtKr01} gave a significant improvement and
reduced the bound to $(\log n)^7 \cdot n^{k}$ clauses for
efficient refutation of $2k$-cnf formulas. This was later improved
by \cite{CojaGoLaSc03}, \cite{FeigeOf04} that showed how to
efficiently refute a random $2k$-cnf instances with at least
$cn^{k}$ clauses. The basic approach for refutation of $2k$-cnf
formulas was later extended in
\cite{FriedmanGoKr01},\cite{GoerdtLa03},\cite{FeigeOf04} to handle
also random 3CNF formulas with $n^{3/2+\e}, \text{poly}(\log n)
\cdot n^{3/2}, cn^{3/2}$ clauses respectively. Our current result
gives a somewhat weak indication that spectral methods can not
break the $n^{3/2-o(1)}$ barrier.

Further motivation for studying efficient refutability of random
3CNF formulas is given in \cite{Feige02}. There it is shown that if
there is no polynomial time refutation heuristic that works for most
3CNF formulas with $cn$ clauses (where $c$ is an arbitrarily large
constant) then certain combinatorial optimization problems (like
minimum graph bisection, the dense $k$-subgraph, and others) have no
polynomial time approximation schemes. It is an open question
whether it is NP-hard to approximate these problems arbitrarily
well, though further evidence that these problems are indeed hard to
approximate is given in~\cite{khot}.

The algorithm considered in the current paper for refuting $\phi$
by computing $\vartheta(G_{\phi})$ was presented
in~\cite{Feige02}. There is was shown that when $m < n^{2-o(1)}$,
almost surely $\vartheta(G_{\phi}) \geq (1 -o(1))m$. Our current
work overcomes a difficulty that prevented the approach
of~\cite{Feige02} to show that $\vartheta(G_{\phi}) = m$, not even
for formulas $\phi$ with a linear number of clauses. The
difficulty was the existence of pairs of clauses that share two
variables.

Related algorithms for refuting CNF formulas were analysed
in~\cite{BGHMP,AAT}. There the authors considered a certain linear
programming relaxation of the satisfiability problem, and
successive tightenings of this relaxation via the operators of
Lovasz and Schrijver. The authors of \cite{AAT} show that in order
to refute a random 3CNF formula with $cn$ clauses (where $c$ is a
sufficiently large constant) one has to apply $\Omega(n)$ rounds
of the Lovasz-Schrijver operator to the initial relaxation. Our
results deal only with the Lovasz $\vartheta$ function which lies
at the lowest level of the Lovasz-Schrijver hierarchy (for maximum
independent set relaxation). In this respect, the results in
\cite{AAT} are stronger than ours. However, we believe that our
results are of independent interest. (In fact, they were obtained
independently of and roughly concurrently with the results of
\cite{AAT}.) One superficial difference is that we consider denser
3CNF formulas. This difference is only superficial, because also
the results of \cite{AAT} extend to denser formulas, by limiting
them to the lower levels of the Lovasz-Schrijver hierarchy. A more
substantial difference is that the staring point of \cite{AAT},
which is a linear program relaxation of 3CNF, is different from
ours. We first apply a reduction to the 3CNF formula, inducing a
graph, and only then apply the Lovasz $\vartheta$ function to the
induced graph. It is not obvious (at least for us) what is the
minimal $i$ for which the $i$-th relaxation used in \cite{AAT} is
stronger than the relaxation we use (such $i$ exists since the
$n$-th relaxation always returns the correct answer). And finally,
there are differences between our proof techniques and those of
\cite{AAT}. We present a solution to the vector formulation of the
$\vartheta$ function, whereas \cite{AAT} present a solution to the
matrix formulation of their relaxation.

\section{Results}
Instead of working with $G_{\phi}$ we work with an induced
subgraph of $G_{\phi}$ that is derived from $G_{\phi}$ by
retaining in each clause cloud only the vertices corresponding to
satisfying 3XOR assignments of the clause. Namely, for each clause
we keep those four vertices that are associated with assignments
that satisfy an odd number of literals in the clause. We call this
subgraph $G_{\phi}^{3xor}$. Since $G_{\phi}^{xor}$ is an induced
subgraph of $G_{\phi}$ it follows (by known monotonicity
properties of the theta function) that $\vartheta(G_{\phi}^{xor})
\leq \vartheta(G_{\phi})$. We show that when $m \leq n^{3/2
-o(1)}$ w.h.p. $\vartheta(G_{\phi}^{xor}) =m$, which by the above
discussion implies that also $\vartheta(G_{\phi}) = m$.

\begin{thm}\label{thm: main}
Let $\phi$ be a random 3CNF formula with $m =o (n^{\frac{3}{2} -
\frac{22 \log \log n}{\log n}})$ clauses and $n$ variables. With
high probability $\vartheta(G_{\phi}^{xor}) = m$.
\end{thm}

\begin{corollary}
Let $\phi$ be a random 3CNF formula with $m =o (n^{\frac{3}{2} -
\frac{22 \log \log n}{\log n}})$ clauses and $n$ variables. With
high probability $\vartheta(G_{\phi}) = m$.
\end{corollary}

For $G_{\phi}^{xor}$ our results are nearly optimal in terms of the
density of the underlying 3CNF formula $\phi$.

\begin{prop}\label{thm: refutation for dense}
Let $\phi$ be a random 3CNF formula with $m \geq cn^{3/2}$ clauses
and $n$ variables, where $c$ is a sufficiently large constant. With
high probability $\vartheta(G_{\phi}^{xor}) < m$.
\end{prop}

We suspect that when $m \geq cn^{3/2}$ then also
$\vartheta(G_{\phi}) < m$, although we did not prove it (when $m
\geq cn^{3/2}$ there are other refutation methods that succeed,
see \cite{FeigeOf04} for details).

For convenience, from now on we will refer to the
$\vartheta(G_{\phi}^{xor})$ also as $SDP(\phi)$. We prove Theorem
\ref{thm: main} in two steps. First we introduce a simple refutation
proof system that we call narrow Gauss Elimination $3$ (in short
GE3) and prove that it is stronger then $SDP(\phi)$, i.e. if $\phi$
cannot be refuted by GE3 then $SDP(\phi) =m$. We then show that a
random 3XOR formula with $m =o (n^{\frac{3}{2} - \frac{22 \log \log
n}{\log n}})$ clauses almost surely cannot be refuted by GE3.

\begin{definition}
The GE3 proof system works as follows. It receives as input a
system of linear equations modulo~2, where every equation has at
most three literals.
It succeeds in refuting the system of linear equations if it manages
to derive the equation $0 = 1$. A new equation can be derived only
if it contains at most three variables, and it is the result of
adding exactly two existing equations and simplifying the result mod
$2$. By simplifying modulo~2 we mean that $1 \pm 1 = 0$, $x_i \pm
x_i = 0$ and $x_i \pm \bar{x}_i = 1$, for every variable~$i$.
\end{definition}

To clarify the derivation rule of GE3, consider the following
three linear equations: $x_1 + x_2 + x_3 = 1$, $x_1 + x_4 + x_5 =
1$ and $x_2 + x_4 + x_6 = 1$. No new equation can be derived by
the GE3 proof system, because adding any two equations produces an
equation with four variables. In particular, also the equation
$x_3 + x_5 + x_6 = 1$ cannot be derived, even though it contains
only three variables and is implied by the original equations (by
adding the three of them).

Observe that if an equation $e_1$ containing only two variables is
derived in GE3 (say, $x_1 + x_2 = 0$), then in every other
equation $e_2$ we can use GE3 to replace the occurrence of one of
the variables by the other, by adding $e_1$ and $e_2$.

The proof of Theorem~\ref{thm: main} is an immediate consequence
of the following two lemmas.

\begin{lemma}\label{lemma: GE3 reduction}
Let $\phi$ be any formula with $m$ clauses. If $\phi$ cannot be
refuted by $GE3$ then $SDP(\phi) = m$.
\end{lemma}

\begin{lemma}\label{lemma: GE3 is weak}
Let $\phi$ be a random 3XOR formula with $n$ variables and $m =o
(n^{\frac{3}{2} - \frac{22 \log \log n}{\log n}})$ clauses. With
high probability GE3 cannot refute $\phi$.
\end{lemma}

\section{SDP formulation of the $\vartheta$ function}\label{sec: SDP
theta}

For each vertex $i$ we assign a vector $v_i$. There is also a
special vector $v_0$. The semidefinite program is:
\begin{align}
&\max \sum_{i=1}^{n} \inprod{v_0}{v_i} &\text{subject to:} \nonumber \\
& &\inprod{v_0}{v_0} = 1\\
&\text{for every $i\geq 1$:}  &\inprod{v_i}{v_i} = \inprod{v_i}{v_0}\label{eqn: 2}\\
&\text{for every pair $i,j$:} &\inprod{v_i}{v_j} \geq 0 \label{eqn: 3}\\
&\text{for any edge $(i,j)$:} &\inprod{v_i}{v_j} = 0 \label{eqn: 4}
\end{align}

Instantiating the above semi-definite program for the graph
$G_{\phi}^{xor}$ we derive the following semi-definite program, in
which for clause $i$ there are $4$ assignment vectors $v_{i}^j$,
one for every assignment of its three variables that satisfies an
odd number of literals in the clause.
\begin{align}
&\max \sum_{\substack{i=1..m,\\j=1..4}} \inprod{v_0}{v_i^j} &\text{subject to:} \nonumber \\
& &\inprod{v_0}{v_0} = 1\\
&\text{for every vector:}  &\inprod{v_i^j}{v_i^j} = \inprod{v_i^j}{v_0}\label{eqn: 6}\\
&\text{for every pair of vectors:} &\inprod{v_i^j}{v_k^l} \geq 0 \label{eqn: 7}\\
&\text{for every pair of contradicting vectors:}
&\inprod{v_i^j}{v_k^l} = 0 \label{eqn: 8}
\end{align}

\noindent (A pair of vectors is contradicting if there is some
variable that the assignment associated with one of the vectors
sets to true, and the assignment associated with the other vector
assigns to false.)

The value of the second semi-definite program is at most $m$
because every clause cloud forms a clique. As the following known
Lemma shows, the contribution of a clique to the objective
function is at most $1$.

\begin{lemma}\label{lemma: theta clique cover}
Let $v_0$ be a unit vector and let $v_1,v_2,v_3,v_4$ be orthogonal
vectors, such that $\inprod{v_i}{v_i} = \inprod{v_i}{v_0}$ for all
$i$. Then $\sum_{i=1}^4 \inprod{v_0}{v_i} \leq 1$.
\end{lemma}

\begin{proof}
Since $v_0$ is a unit vector and $v_1,v_2,v_3,v_4$ are orthogonal,
it holds that $\sum_{i=1}^4 \inprod{v_0}{\frac{v_i}{\norm{}{v_i}}}^2
\leq 1$. It thus follows that
\begin{align*}
& \sum_{i=1}^4 \inprod{v_0}{v_i} =\sum_{i=1}^4 \norm{}{v_i}
\inprod{v_0}{\frac{v_i}{\norm{}{v_i}}}  = \sum_{i=1}^4
\inprod{v_0}{\frac{v_i}{\norm{}{v_i}}}^2 \leq 1,
\end{align*}
where the last equality follows from $\norm{}{v_i}^2 =
\inprod{v_0}{v_i}$.
\end{proof}
Note that Lemma \ref{lemma: theta clique cover} implies that for any
graph $G$, if the vertices of $G$ can be covered by $p$ cliques,
then $\vartheta(G) \leq p$.

\section{Proofs}

We will use the SDP formulation of the $\vartheta$ function as
appears in Section \ref{sec: SDP theta}.
\begin{proof}[Proof of lemma \ref{lemma: GE3 reduction}]
Apply the derivation rule of the GE3 system as long as new
equations are generated by it. Since the number of possible
equations with at most three variables is $O(n^3)$, then this
procedure must end. Assume that the equation $0 = 1$ could not be
derived. Hence we are left with equations containing one variable
(meaning that the value of this variable must be fixed to a
constant), two variables (meaning that their values must be
identical, or sum up to 1, depending on the free constant in the
equation), or three variables. The information that GE3 derives
about $\phi$ allows us to partition all literals into equivalence
classes of the form:
\begin{align}
S_1:~~ &x_1 = x_{18} = \ldots = \bar{x}_9 &(\bar{S}_1:~~ \bar{x}_1 = \bar{x}_{18} = \ldots = x_9) \nonumber\\
S_2:~~ &x_4 = x_{20}  = \ldots = x_5 &(\bar{S}_2:~~\bar{x}_4 = \bar{x}_{20}  = \ldots = \bar{x}_5) \nonumber\\
&~~~~. &.~~~~~~~~~  \\
&~~~~. &.~~~~~~~~~  \nonumber\\
S_9:~~ &\bar{x}_2 = x_{21}= \dots = x_{30} &(\bar{S}_9:~~ x_2= \bar{x}_{21}= \dots = \bar{x}_{30} )\nonumber \\
&~~~~. &.~~~~~~~~~  \nonumber\\
 S_l:~~ &1=x_6 = \bar{x}_{11}= \ldots x_8
&(\bar{S}_l:~~0=\bar{x}_6 = x_{11}= \ldots \bar{x}_8) \nonumber\\
\nonumber
\end{align}
Notice that each equivalence class $S_i$ has a "mirror" part
$\bar{S}_i$; we think of these two parts as one class. A class
might contain only one variable. We call a variable \emph{fixed}
if it belongs either to $S_l$ or to the mirror of $S_l$.  Other
variables are called \emph{free}. Similarly, except $S_l$ which is
fixed, all other classes are free. A variable is fixed if and only
if the GE3 refutation system can derive a clause containing only
this variable (equal to a constant). Two free variables belong to
the same class if and only if the GE3 system can derive a clause
containing only these two variables.

Each original clause of $\phi$ is of one of the following types:
\begin{enumerate}
\item It contains three free variables, each of them has distinct
equivalence class.

\item It contains one fixed variable and two
free variables from the same equivalence class.

\item It contains
three fixed variables.
\end{enumerate}
We now explain why the above three types cover all clauses. If a
clause has no fixed variable then its variables must be from
distinct classes (type 1), as otherwise two of them will cancel
out and cause the other variable to be fixed. If a clause has
exactly one fixed variable then the other two belong to the same
class and they are free (type 2). A clause cannot have exactly two
fixed variables as the remaining variable will be also fixed (thus
the remaining case is type 3).

We will now give values to the vectors corresponding to all clauses.
These vectors will satisfy the SDP constraints and will also give a
value of $m$. An assignment for a clause that contradicts the
information gathered by GE3 is called \emph{illegal}; otherwise it
is \emph{legal}. For example, for the equivalence classes given
above, an assignment such as $x_{11} = 1$ is illegal because it
contradicts $S_l$, also an assignment such as $(x_1,x_9,x_{11}) =
(1,1,0)$ is illegal because it contradicts $S_1$. We will use the
following guidelines:
\begin{itemize}
\item Each vector has $l$ coordinates, numbered from 0 to $l-1$.
For $1 \le i \le l-1$, coordinate $i$ will correspond to free
class $i$

\item A clause vector that corresponds to an illegal assignment
will be set to the zero vector $\vec{0}$. For a clause of type (1)
the clause cloud will have four assignments with non zero vectors,
for a clause of type (2) there will be two assignments, and for a
clause of type (3) there will be one assignment.

\item Let $c$ be a clause that has $i$ different free classes ($i \in
\{0,1,3\}$). The vectors corresponding to legal assignments of $c$
will have exactly $1 + i$ non zero entries. The only non-zero
coordinates are $0$ and the coordinates corresponding to the
indices of the free classes.
\end{itemize}

Notice that the second bullet can be interpreted as removing from
$G_{\phi}^{xor}$ all the vertices corresponding to illegal
assignments. Thus from now we will assume that such vertices are
indeed removed from $G_{\phi}^{xor}$. To simplify the notation in
the remainder of the proof, we do the following. With each
subclass $S_i$ we associate a literal $s_i$ (and with $\bar{S}_i$
we associate $\bar{s}_i$). We translate each clause
$c=(\bar{x}_9,x_2,x_5)$ into a new clause
$\tilde{c}=(s_1,\bar{s}_9, s_2)$ by replacing each literal $x_i$
of $c$ with the literal corresponding the unique subclass which
contains $x_i$. Note that the subclass literal replacing the
literal $x_i$ may have polarity opposite to $x_i$ (if for example
$x_i \in \bar{S}_i$). The new induced formula $\tilde{\phi}$ may
contain some clauses with multiplicity $>1$ as well as clauses in
which some variable appears more than once (e.g. $(s_1,s_1,s_8)$).
We will now define a homomorphism $f$ from $G_\phi^{xor}$ to
$G_{\tilde{\phi}}^{xor}$, which implies that
$\vartheta(G_\phi^{xor}) \geq \vartheta(G_{\tilde{\phi}}^{xor})$
(a homomorphism $f:G \rightarrow H$ maps the vertices of $G$ into
the vertices of $H$ while preserving the edge relation, i.e. if
$(u,v) \in E(G)$ then $(f(u),f(v)) \in E(H)$). Recall that each
clause $c = (\bar{x}_9,x_2,x_5)$ of $\phi$ has a unique
corresponding clause $\tilde{c} = (s_1,\bar{s}_9, s_2)$ of
$\tilde{\phi}$ (although other copies of $(s_1,\bar{s}_9, s_2)$
may exist in $\tilde{\phi}$). The map $f$ is defined only for
legal satisfying assignments of $\phi$ (we already removed from
$G_{\phi}^{xor}$ all the non legal assignments). $f$ maps the
vertices (assignments) in the clause cloud of $c$ to vertices
(assignments) in the clause cloud of
$\tilde{c}$ as follows:\\
for a \textbf{legal} satisfying assignment of $c$, say $(\bar{x}_9
,x_2,x_5)= (1,1,1)$, we replace each literal $x_i$ with its
corresponding class literal and leave the values as is. For example
if $\bar{x}_9 \in S_1, x_2 \in \bar{S}_9, x_5 \in S_2$ then $f$ maps
the assignment $(\bar{x}_9 ,x_2,x_5) = (1,1,1)$ into
$(s_1,\bar{s}_9, s_2) = (1,1,1)$. It is not hard to see that $f$
maps a legal satisfying assignment for $c$ into an assignment for
$\tilde{c}$ that is both satisfying and noncontradictory (meaning
for example that it will not result in one occurrence of $s_1$ being
set to~0 and the other being set to~1). The assignment $f$ returns
must be non contradictory as otherwise $\phi$ can be refuted by GE3.
Note that GE3 can not refute $\tilde{\phi}$ nor can it derive an
equation like $s_i = s_j$, for $i \neq j$. From here on we show a
SDP solution to $G_{\tilde{\phi}}^{xor}$.

The vector $v_0$ is set to be $(1,0,\ldots,0)$. The remaining vector
assignments are as follows, divided by the clause types:
\begin{enumerate}
\item Type (1), three free distinct classes. Assume the clause is
$\tilde{c} = (\bar{s}_{1},s_2,s_4)$. The vector assignments will
be:
\begin{align*}
&v^{\tilde{c}}_{(\bar{s}_{1},s_2,s_4) =(1,1,1)} = (\nicefrac{1}{4},~
- \nicefrac{1}{4},~~~~~\nicefrac{1}{4},~~~0,~~~~\nicefrac{1}{4},~~~0,~\ldots,0)\\
&v^{\tilde{c}}_{(\bar{s}_{1},s_2,s_4)=(1,0,0)} = (\nicefrac{1}{4},~
- \nicefrac{1}{4},~-\nicefrac{1}{4},~~~0,~-\nicefrac{1}{4},~~~0,~\ldots,0)\\
&v^{\tilde{c}}_{(\bar{s}_{1},s_2,s_4)=(0,1,0)} = (\nicefrac{1}{4},
~~~~~\nicefrac{1}{4},~~~~~\nicefrac{1}{4},~~~0,~-\nicefrac{1}{4},~~~0,~\ldots,0)\\
&v^{\tilde{c}}_{(\bar{s}_{1},s_2,s_4)=(0,0,1)} = (\nicefrac{1}{4},
~~~~~\nicefrac{1}{4},~-\nicefrac{1}{4},~~~0,~~~~~\nicefrac{1}{4},~~~0,~\ldots,0)
\end{align*}
\item Type (2), one fixed class and two occurrences of some free
class. Hence the equation has exactly two satisfying assignments.
One assignment would get a vector that has $1/2$ in its 0
coordinate and  $1/2$ on the coordinate corresponding to the free
class, and the other would get a vector that has $1/2$ in its 0
coordinate and $-1/2$ on the coordinate corresponding to the free
class. For example, for the clause $\tilde{c} = (s_l,s_2,s_2)$ the
vectors would be:
\begin{align*}
&v^{\tilde{c}}_{(s_l,s_2,s_2) = (1,1,1)} = (\nicefrac{1}{2} ,~ 0 ,~~~~ \nicefrac{1}{2},~~~ 0,~\ldots,0)\\
&v^{\tilde{c}}_{(s_l,s_2,s_2) = (1,0,0)} = (\nicefrac{1}{2},~ 0,~
-\nicefrac{1}{2} ,~~~ 0,~\ldots,0)
\end{align*}

\item Type (3), three fixed classes. Assume $\tilde{c} =
(s_l,\bar{s}_l,\bar{s}_l)$. In this case the only non-zero vector
is:
\begin{align*}
&v^{\tilde{c}}_{(s_l,\bar{s}_l,\bar{s}_l) = (1,0,0)} = (1,0
,0,0,\ldots,0)
\end{align*}
\end{enumerate}

We next show that the above vector configuration is a valid
solution of the $\vartheta$ function of $G_{\tilde{\phi}}^{xor}$
(it is easy to see that the above solution has value of $m$).
Constraints of type \eqref{eqn: 2} hold because of the special
form of non-zero vectors. The fact that constraints of type
\eqref{eqn: 3} hold will be implicit in our proof that constraints
of type \eqref{eqn: 4} hold, and is omitted. Hence we will only
consider now constraints of type \eqref{eqn: 4}.

Observe first that within every clause cloud constraints of type
\eqref{eqn: 4} hold. Hence it remains to check \eqref{eqn: 4} for
pairs of different clauses that have an $s$ variable in common.
Let $\tilde{c}_1,\tilde{c}_2$ be two clauses that intersect. We
continue by case analysis according to the number of distinct $s$
variables shared by $\tilde{c}_1,\tilde{c}_2$.
\begin{enumerate}
\item Three distinct variables are shared: since GE3 did not deduce $0=1$ the
clauses are identical and \eqref{eqn: 4} (and \eqref{eqn: 3}) hold
from the fact that it holds for each cloud separately.

\item Two distinct variables are shared: using GE3 we deduce that also the
third variable is shared and this case was already handled.

\item Exactly one variables is shared: for simplicity, assume that
each of the clauses contain $3$ different variables and say $s_i$
is the shared variable. The only two indices that contribute to
the inner product sum are $0$ and (possibly) $i$. If $s_i$ is
fixed the assignments cannot be contradictory and the sum is
strictly positive (only coordinate $0$ contribute to the sum).
Assume that $s_i$ is free. Consider the case in which in each
clause the other two literals are also free. If the vectors are of
contradicting assignments the sum will be $(1/4)(1/4) +
(-1/4)(1/4)$ (or $(1/4)(1/4) + (1/4)(-1/4)$). If the vectors are
not of contradicting assignments, the sum is strictly positive.

Note that also in the other cases where one of the clauses
contains only one or two different $s$ variables, a similar
argument works.

\end{enumerate}
\end{proof}

\begin{proof}[Proof of Lemma \ref{lemma: GE3 is weak}]
We follow the line of proof given at \cite{Ben-SassonWi01} with some
simplifications that can be applied in our case. We use the
following definitions from \cite{Ben-SassonWi01}. Let $A,B$ be any
two formulas. $A \models B$ if every satisfying assignment for $A$
is a satisfying assignment for $B$, or equivalently,  every
non-satisfying assignment of $B$ is also a non-satisfying assignment
of $A$. Let $\phi$ be a formula (collection of clauses) and let $C$
be any clause. We use $\mu_{\phi}(C)$ to denote the minimum size
subformula of $\phi$ that implies $C$, i.e. $\mu_{\phi}(C) \eqdef
\min_{ \phi' \subseteq \phi} |\{\phi' \models C\}|$. As $\phi$ is
known from the context (and fixed) we use $\mu(C)$ instead of
$\mu_{\phi}(C)$. The function $\mu$ is sub-additive, meaning that if
$A,B \models C$ then $\mu(C) \leq \mu(A) + \mu(B)$. We use $0$ to
denote a contradiction (the empty clause).

A simple counting argument shows that any subformula of $\phi$ of
size smaller than $k \eqdef \frac{\log n}{4\log \log n}$ is
satisfiable; see Lemma \ref{lemma: random formula properties}. Thus,
$\mu(0) \geq k$. From the sub-additivity of $\mu$, it follows that
any GE3 proof of $0$ contains some clause $C$ for which $\frac{k}{3}
\leq \mu(C) \leq \frac{2k}{3}$ (the explanation is as follows. The
derivation of $0$ can be described by a tree in which every leaf has
a label that equals to some clause of $\phi$ and the root has a
label that equals $0$. For each leaf label, say $A$, it holds that
$\mu(A)=1$ and for the root label $0$ it holds $\mu(0) \geq k$). In
other words, the minimal subformula $E'$ that implies $C$ is of size
in $[\frac{k}{3},\frac{2k}{3}]$. The subformula $E'$ (as any other
subformula of $\phi$, whose size in $[\frac{k}{3},\frac{2k}{3}]$,
see Lemma \ref{lemma: random formula properties}) has at least $4$
\emph{special variables}, each of them appears in exactly one clause
of $E'$. We show in the next paragraph that each of these $4$
special variables must be in $C$. This implies that $C$ cannot be
derived in GE3, contradicting the assumption that GE3 refutes
$\phi$.

Let $x$ be a special variable that belongs to some clause $f$ of
$E'$ (and not to any other clause in $E'$). From the minimality of
$E'$, there exists an assignment $\alpha$ such that $f(\alpha) =
C(\alpha) = 0$ but for any other clause $g \in E'$ it holds that
$g(\alpha)=1$ (as otherwise $E' \setminus \{f\} \models C$). By
contradiction, assume that $x \not \in C$. Changing the value of
$\alpha$ only on $x$ leaves $C$ unsatisfied. Yet, $f$ becomes
satisfied and any other clause of $E'$ remains satisfied because
$x$ appears only on $f$. We deduce that after changing $\alpha$
only on $x$ the subformula $E'$ becomes satisfied while $C$ is
not, this is a contradiction to $E' \models C$.

\end{proof}

\begin{lemma} \label{lemma: random formula properties}
Let $\phi$ be a random formula with $m = o\left( n^{\frac{3}{2} -
\frac{22 \log \log n}{\log n}} \right)$ clauses. Set $k = \frac{\log
n}{4\log \log n}$. With high probability the following properties
hold.
\begin{enumerate}
\item Any subformula of $\phi$ of size $k$ is satisfiable.
\item Any subformula $E' \subset \phi$, whose size is in $[\frac{k}{3},\frac{2k}{3}]$, has at least $4$ variables,
each of them belongs to exactly one clause of $E'$.
\end{enumerate}
\end{lemma}

\begin{proof}
We show that any small subformula $\phi'$ is satisfiable by showing
that in any such small subformula, the number of variables is at
least the number of clauses. By Hall's marriage theorem, in any such
subformula $\phi'$ there is a matching from the variables to the
clauses that covers all the clauses, which implies that $\phi'$ is
satisfiable. We now analyse the first event (proving part 1 of the
lemma). Consider $k$ clauses chosen at random. The probability that
they contain less than $k$ different variables is bounded by the
probability of the following event: when throwing $3k$ balls into
$n$ bins, the set of non empty bins is $<k$.  Thus the probability
for the first event is at most
\begin{align*}
& {m \choose k} \sum_{i=1}^{k-1}  {n \choose i} \left(
\frac{i}{n}\right)^{3k} \leq 2 \left( \frac{me}{k} \right)^k
 \left( \frac{ne}{k-1} \right)^{k-1} \left(
\frac{k-1}{n}\right)^{3k} \\
& \leq 2\frac{e^{2k-1}(k-1)^{k+1}}{n} \left(
\frac{m}{n^2}\right)^{k} \leq o(1) \left( \frac{m}{n^2}\right)^{k}.
\end{align*}
(the first inequality is because the sum is geometric with ratio
$\geq \frac{en}{k}$, the last inequality holds for $k = \frac{\log
n}{4\log \log n}$).

We now bound the probability of the second event (part $2$ of the
lemma). Fix $l$ to be in the interval $[\frac{k}{3},\frac{2k}{3}]$.
Consider $l$ clauses chosen at random. The probability that they
contain less than $4$ special variables equals the probability of
the following event. When throwing $l$ triplets of balls into $n$
bins (where each triplet of balls choose three different bins) there
are less than $4$ bins that contain exactly one ball. Notice that if
the balls fall into more than $3(l+1)/2$ bins, there must be at
least $4$ bins that contain exactly one ball. The probability is
thus bounded by
\begin{align*}
{m \choose l} \sum_{i=1}^{3(l+1)/2}  {n \choose i} \left( \frac{{i
\choose 3}}{{n \choose 3}}\right)^{l} &\leq 2 \left( \frac{me}{l}
\right)^l  \left( \frac{ne}{3(l+1)/2} \right)^{3(l+1)/2} \left( 1.01
\frac{3(l+1)/2}{n}\right)^{3l} \\
& \leq l^{4l} \left( \frac{m}{n^{\frac{3}{2}(1 -
\frac{1}{l})}}\right)^{l} \leq \left( \frac{ml^4}{n^{\frac{3}{2}(1 -
\frac{1}{l})}}\right)^{l} .
\end{align*}
To cover all possible values of $l \in [\frac{k}{3},\frac{2k}{3}]$
we multiply the last term by $k$. The induced bound is $o(1)$ for $m
= o\left( n^{\frac{3}{2} - \frac{22 \log \log n}{\log n}} \right)$.

\end{proof}

\begin{proof}[Proof of Proposition \ref{thm: refutation for dense}]
A simple probabilistic argument shows that if $c$ is large enough,
$\phi$ is likely to contains four clauses of the following form (see
Lemma \ref{lemma: four clauses}):
\begin{align*}
&c_1 = (x_1,x_2,x_3) & c_3 = (x_5,x_6,x_3)\\
&c_2 = (x_1,x_2,x_4) & c_4 = (x_5,x_6,\bar{x}_4)
\end{align*}
The above four clauses are contradictory (summing all of them give
$1=0$ modulus $2$).

The $\vartheta$ function of the graph induced only by these $4$
clauses has a value of $\approx 3.4142 < 4$. This bound was
experimentally derived by running a semi-definite programming
package on Matlab. The adjacency matrix we used is:
\[
\begin{MAT}(e,3pt,3pt)[3pt]{cccccccccccccccc}
0 &1 &1 &1 &0 &1 &1 &1 &0 &1 &1 &0 &0 &0 &0 &0\\
1 &0 &1 &1 &1 &0 &1 &1 &1 &0 &0 &1 &0 &0 &0 &0\\
1 &1 &0 &1 &1 &1 &0 &1 &1 &0 &0 &1 &0 &0 &0 &0\\
1 &1 &1 &0 &1 &1 &1 &0 &0 &1 &1 &0 &0 &0 &0 &0\\
0 &1 &1 &1 &0 &1 &1 &1 &0 &0 &0 &0 &1 &0 &0 &1\\
1 &0 &1 &1 &1 &0 &1 &1 &0 &0 &0 &0 &0 &1 &1 &0\\
1 &1 &0 &1 &1 &1 &0 &1 &0 &0 &0 &0 &0 &1 &1 &0\\
1 &1 &1 &0 &1 &1 &1 &0 &0 &0 &0 &0 &1 &0 &0 &1\\
0 &1 &1 &0 &0 &0 &0 &0 &0 &1 &1 &1 &0 &1 &1 &1\\
1 &0 &0 &1 &0 &0 &0 &0 &1 &0 &1 &1 &1 &0 &1 &1\\
1 &0 &0 &1 &0 &0 &0 &0 &1 &1 &0 &1 &1 &1 &0 &1\\
0 &1 &1 &0 &0 &0 &0 &0 &1 &1 &1 &0 &1 &1 &1 &0\\
0 &0 &0 &0 &1 &0 &0 &1 &0 &1 &1 &1 &0 &1 &1 &1\\
0 &0 &0 &0 &0 &1 &1 &0 &1 &0 &1 &1 &1 &0 &1 &1\\
0 &0 &0 &0 &0 &1 &1 &0 &1 &1 &0 &1 &1 &1 &0 &1\\
0 &0 &0 &0 &1 &0 &0 &1 &1 &1 &1 &0 &1 &1 &1 &0\\
\end{MAT}
\]
vertices $1,2,3,4$ correspond to $c_1$ ,vertices $5,6,7,8$
correspond to clause $c_2$, vertices $9,10,11,12$ correspond to
clause $c_3$ and vertices $13,14,15,16$ correspond to clause $c_4$:

\begin{tabbing}
aaa \= aaa \= aaa \= aaa \= aaaaa \= aaa \= aaa \= aaa \= aaa \=
aaa \kill {}
\>      \> $x_1$ \> $x_2$ \> $x_3$ \>      \> $x_1$ \> $x_2$ \> $x_4$\\
\> $v_1$ \> 1     \>   1   \>   1 \> $v_5$ \> 1     \>   1   \>   1\\
\> $v_2$ \> 0     \>   1   \>   0 \> $v_6$ \> 0     \>   1   \>   0\\
\> $v_3$ \> 1     \>   0   \>   0 \> $v_7$ \> 1     \>   0   \>   0\\
\> $v_4$ \> 0     \>   0   \>   1 \> $v_8$ \> 0     \>   0   \>   1\\
\end{tabbing}

\begin{tabbing}
aaa \= aaa \= aaa \= aaa \= aaaaa \= aaa \= aaa \= aaa \= aaa \=
aaa \kill {}
\>      \> $x_5$ \> $x_6$ \> $x_3$ \>      \> $x_5$ \> $x_6$ \> $\bar{x}_4$\\
\> $v_9$ \> 1     \>   1   \>   1 \> $v_{13}$ \> 1     \>   1   \>   1\\
\> $v_{10}$ \> 0     \>   1   \>   0 \> $v_{14}$ \> 0     \>   1   \>   0\\
\> $v_{11}$ \> 1     \>   0   \>   0 \> $v_{15}$ \> 1     \>   0   \>   0\\
\> $v_{12}$ \> 0     \>   0   \>   1 \> $v_{16}$ \> 0     \>   0   \>   1\\
\end{tabbing}

The $\vartheta$ function of $G_{\phi}$ must be smaller than $<m$
as the remaining graph (without the clouds of $c_1,c_2,c_3,c_4$)
can be covered by $m-4$ cliques.

\end{proof}

\begin{lemma}\label{lemma: four clauses}
Let $\phi$ be a random formula with $n$ variables and $m=cn^{3/2}$
random clauses. Almost surely $\phi$ contains four clauses of the
form:
\begin{align*}
&c_1 = (x_1,x_2,x_3) & c_3 = (x_5,x_6,x_3)\\
&c_2 = (x_1,x_2,x_4) & c_4 = (x_5,x_6,\bar{x}_4)
\end{align*}
\end{lemma}
\begin{proof}
We say that $a(n) \sim b(n)$ if $\lim_{n \rightarrow \infty}
\frac{a(n)}{b(n)} =1$. A pair of clauses is said to \emph{match} if
the two clauses share the same first and second literal. The
expected number of matched pairs in $\phi$ is
\begin{align}
{m \choose 2} \frac{1}{2n}~\frac{1}{2n-2} \sim \frac{c^2n^3}{2}
\frac{1}{4n^2} = \frac{c^2n}{8}.
\end{align}
Furthermore, it can be shown that w.h.p. $\phi$ contains $\sim
\frac{c^2n}{8}$ matched pairs such that each clause of $\phi$
participates in at most one pair of matching clauses (a standard
use of the second moment, see for example \cite{FeigeOf04} for a
proof).
Assume we have
$\sim \frac{c^2n}{8}$ matched pairs. For any such pair the third
literal in each of them is still random. Fix two matched pairs
$c_1,c_2$ and $c_3,c_4$. With probability $\sim \frac{1}{4n^2}$ the
third literal of $c_1$ and $c_3$ is the same and the third literal
of $c_2$ is opposite from the third literal of $c_4$. It thus
follows that the expected number of two pairs of the form
\begin{align*}
&c_1 = (x_1,x_2,x_3) & c_3 = (x_5,x_6,x_3)\\
&c_2 = (x_1,x_2,x_4) & c_4 = (x_5,x_6,\bar{x}_4),
\end{align*}
is
\begin{align}
\sim \frac{1}{2} \left(\frac{c^2n}{8} \right)^2 \frac{1}{4n^2} \sim
\frac{c^4}{8^3}.
\end{align}
Using standard techniques (such as the second moment), it can be
shown that almost surely $\phi$ contains four clauses of this form.
Details are omitted.
\end{proof}

\section*{Acknowledgements}
This work was supported in part by a grant from the German-Israeli
Foundation for Scientific Research and Development (G.I.F.).


\begin{thebibliography}{99}

\bibitem{AAT}
M.~Alekhnovich, S.~Arora and I.~Tourlakis. Towards strong
nonapproximability results in the Lovasz-Schrijver hierarchy. STOC
2005, 294--303.

\bibitem{BGHMP}
J.~Buresh-Oppenheim, N.~Galesi, S.~Hoory, A.~Magen and T.~Pitassi.
Rank bounds and integrality gaps for cutting plane procedures.
FOCS 2003.

\bibitem{Ben-SassonWi01}
E. Ben-Sasson and A. Wigderson.
\newblock Short proofs are narrow -—resolution made simple.
\newblock J. ACM, 48(2):149--169, 2001.

\bibitem{ChvatalSz88}
V.~Chvatal and E.~Szemeredi.
\newblock Many hard examples for resolution.
\newblock J. ACM, 35(4):759--768, Oct 1988.

\bibitem{CojaGoLaSc03}
A.~Coja-Oghlan, A.~Goerdt, A.~Lanka, and F.~Schadlich.
\newblock Certifying unsatisfiability of random 2k-sat formulas using
  approximation techniques.
\newblock FCT 2003, 15--26.

\bibitem{Feige02}
U. Feige.
\newblock Relations between average case complexity and approximation
complexity.
\newblock STOC 2002, 534--543.

\bibitem{FeigeOf04}
U. Feige and E. Ofek.
\newblock Easily refutable subformulas of large random 3cnf
formulas.
\newblock ICALP 2004, 519--530.


\bibitem{FriedgutBo99}
E.~Friedgut and J.~Bourgain.
\newblock Sharp thresholds of graph properties, and the k-sat problem.
\newblock J. of the American Mathematical Society, 12(4):1017--1054, 1999.

\bibitem{FriedmanGoKr01}
J.~Friedman, A.~Goerdt, and M.~Krivelevich.
\newblock Recognizing more unsatisfiable random 3-sat instances efficiently.
\newblock Technical report, 2003.


\bibitem{GoerdtKr01}
A.~Goerdt and M.~Krivelevich.
\newblock Efficient recognition of random unsatisfiable k-{SAT} instances by
  spectral methods.
\newblock STACS 2001, 294--304.

\bibitem{GoerdtLa03}
A.~Goerdt and A.~Lanka.
\newblock Recognizing more random 3-sat instances efficiently.
\newblock Manuscript, 2003.

\bibitem{HajiaghayiSo03}
M. Hajiaghayi  and G.B. Sorkin.
\newblock The satisfiability threshold for random 3-SAT is at least 3.52.
\newblock http://arxiv.org/abs/math.CO/0310193, 2003.

\bibitem{JansonStVa00}
S.~Janson, Y.~C. Stamatiou, and M.~Vamvakari.
\newblock Bounding the unsatisfiability threshold of random 3-sat.
\newblock Random Structures and Algorithms, 17(2):103--116, 2000.

\bibitem{KaporisKiLa03}
A.C. Kaporis, L.M. Kirousis, and E.G. Lalas.
\newblock Selecting complementary pairs of literals.
\newblock LICS 2003.

\bibitem{khot}
S.~Khot. Ruling Out PTAS for Graph Min-Bisection, Densest Subgraph
and Bipartite Clique. FOCS 2004, 136--145.

\end{thebibliography}
\end{document}